\documentclass[sigconf]{aamas}  

\usepackage{booktabs}
\usepackage{algorithm}
\usepackage{algorithmic}
\usepackage{booktabs}
\usepackage{wrapfig}
\usepackage{subcaption}
\usepackage{graphicx}
\usepackage{balance}
\setcopyright{ifaamas}  
\acmDOI{}  
\acmISBN{}  
\acmConference[AAMAS'18]{Proc.\@ of the 
17th International Conference on Autonomous 
Agents and Multiagent Systems (AAMAS 2018)}
{July 10--15, 2018}{Stockholm, Sweden}
{M.~Dastani, G.~Sukthankar, E.~Andr\'{e}, S.~Koenig (eds.)}  
\acmYear{2018}  
\copyrightyear{2018}  
\acmPrice{}  

\begin{document}

\title{Human-Interactive Subgoal 
Supervision for Efficient Inverse 
Reinforcement Learning}  



\author{Xinlei Pan}
\affiliation{
  \institution{University of California, Berkeley}
  \city{Berkeley} 
  \state{California, USA} 
  \postcode{94720}
}
\email{xinleipan@berkeley.edu}

\author{Eshed Ohn-Bar}
\affiliation{
  \institution{Carnegie Mellon University}
  \city{Pittsburgh} 
  \state{Pennsylvania, USA} 
  \postcode{15213}
}
\email{eshedohnbar@gmail.com}
\author{Nicholas Rhinehart}
\affiliation{
  \institution{Carnegie Mellon University}
  \city{Pittsburgh} 
  \state{Pennsylvania, USA} 
  \postcode{15213}
}
\email{nrhineha@cs.cmu.edu}
\author{Yan Xu}
\affiliation{
  \institution{Carnegie Mellon University}
  \city{Pittsburgh} 
  \state{Pennsylvania, USA} 
  \postcode{15213}
}
\email{yxu2@andrew.cmu.edu}

\author{Yilin Shen}
\affiliation{%
  \institution{Samsung Research America}
  \city{Mountain View} 
  \state{California, USA} 
  \postcode{94043}
}
\email{yilin.shen@samsung.com}

\author{Kris M. Kitani}
\affiliation{
  \institution{Carnegie Mellon University}
  \city{Pittsburgh} 
  \state{Pennsylvania, USA} 
  \postcode{15213}
}
\email{kkitani@cs.cmu.edu}

\begin{abstract}  
Humans are able to understand and perform complex tasks 
by strategically structuring the tasks into incremental steps
or subgoals. For a robot attempting to learn to perform 
a sequential task with critical subgoal states, such states can provide a natural opportunity for 
interaction with a human expert. This paper 
analyzes the benefit of incorporating a notion of subgoals
into Inverse Reinforcement Learning (IRL) with 
a Human-In-The-Loop (HITL) framework. The learning process 
is interactive, with a human expert first providing input 
in the form of full demonstrations along with some subgoal states. 
These subgoal states define a set of subtasks for the learning 
agent to complete in order to achieve the final goal. The learning agent queries for partial demonstrations corresponding to each subtask as needed when the agent struggles with the subtask. The proposed Human Interactive IRL (HI-IRL) framework
is evaluated on several discrete path-planning tasks. We 
demonstrate that subgoal-based interactive structuring of 
the learning task results in significantly more efficient 
learning, requiring only a fraction of the demonstration 
data needed for learning the underlying reward function 
with the baseline IRL model. 
\end{abstract}

%


\maketitle


\section{Introduction}

\begin{figure}[tb]
    \centering
    \includegraphics[width=0.8\linewidth]{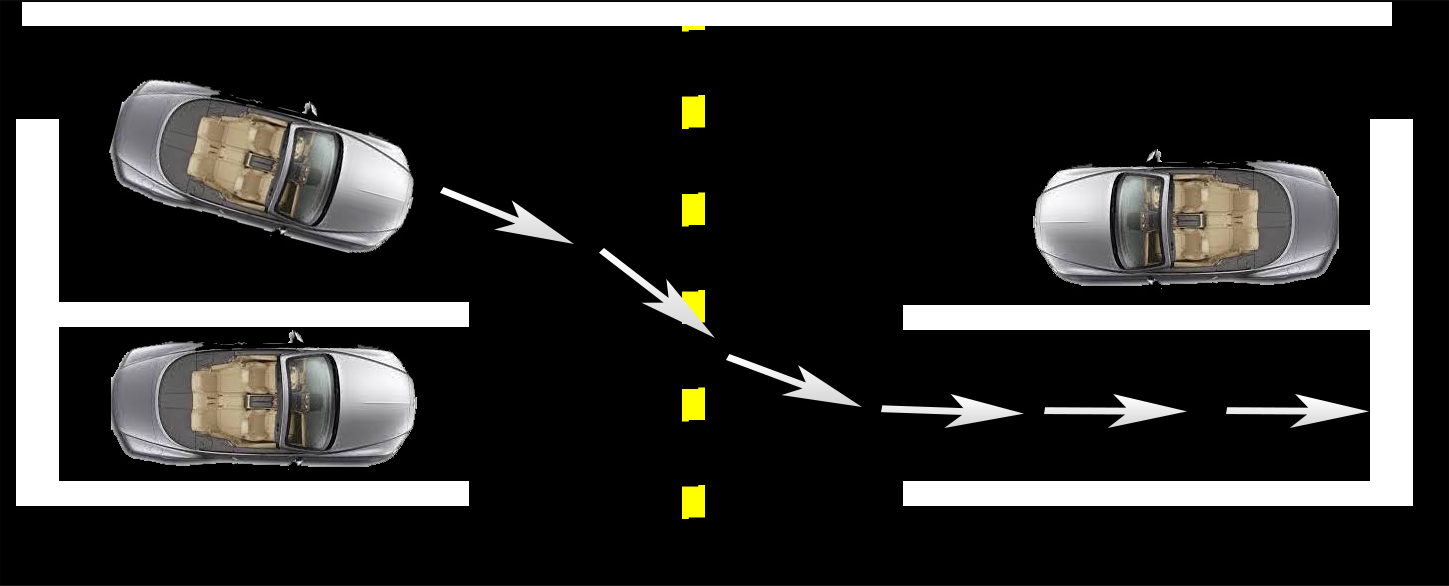}\\
    \caption{
   We develop a framework for training agents that can perform complex sequential tasks with a set of critical subgoals, such as when parking a car. In the example scenario, the car must be first positioned in a certain set of states before being able to continue and complete the goal. By interactively leveraging information regarding subgoal states and subtask demonstration as needed from a human expert, our proposed approach is shown to result in more efficient learning of the underlying reward function. 
    }
    \label{parking_car}
\end{figure}

Teaching robots to perform a sequential, complex task 
is a long-standing research problem in robot learning. 
For instance, consider the task of parking a car into 
a narrow slot as shown in Figure~\ref{parking_car}. 
The autonomous vehicle may be taught to sequentially 
move towards the target across roads while avoiding 
obstacles such as other cars and white lines in the 
environment. One key problem that arises is that while
it can be easy for the car to travel on roads, the car
might struggle locating a specific turning point so 
that it can fit within the narrow parking slot, or 
struggle avoiding hitting other cars when it turns 
around. These issues arise because there are certain 
critical states, namely, subgoal states, that the 
agent \textit{must} visit in order to complete the 
entire task. In this example, the car must turn left 
somewhere before it reaches the empty parking space. 

Leveraging human input is one way to provide information that could 
be helpful for learning agents, like robots, to reach important 
subgoal states. Specifically, a human expert can provide 
demonstrations of possible trajectories to go through these 
critical states for the robot to follow. This type of learning, 
termed broadly as apprenticeship learning 
\cite{abbeel2004apprenticeship,ng2000algorithms}, is a popular approach for 
leveraging human input. 

\begin{figure*}[tb]
    \centering
    \includegraphics[width=0.5\linewidth]{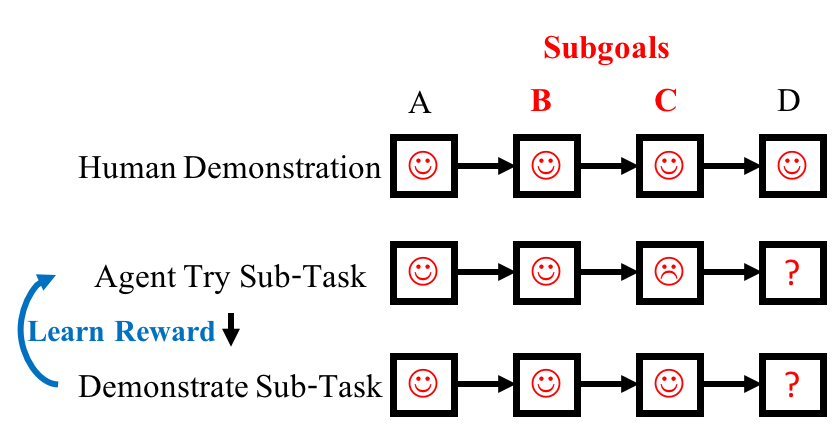}
    \caption{ Diagram of our proposed approach. A human-expert 
    can leverage subgoal states in order to efficiently 
    interact with the learning process. Human will first 
    provide a full demonstration covering the entire task
    (from A to D, these states are landmarks states in the task
    and there are other intermediate states not shown here), 
    and define subgoals (B and C) and subtasks
    (from A to B, B to C and C to D). Then the agent 
    will attempt the human-defined subtasks. Next, the human will only provide subtask demonstration where the agent fails. In this example, the human expert first demonstrates the entire task, and let the agent learn to perform the task. However, the agent may only finish the subtask from A to B (smiley face) but fail 
    on the subtask from B to C (sad face), and stop at C. 
    Then the human expert will demonstrate the subtask (B to C)
    that the agent failed on, and let the agent learn again. This process will repeat until the agent learns to perform the entire task.}
    \label{our_approach}
\end{figure*}

Unfortunately, expert demonstrations might not address all 
of the learning challenges for the following reasons: (1) 
\textit{\textbf{Data Sparsity}} - While an expert can provide 
demonstrations of the entire task, these demonstrations are 
usually collected without considering the learning process 
(\emph{i.e.} the structure of the task and difficulties of 
individual parts). Due to lack of enough demonstrations on 
some critical states, figuring out the way to go through them 
can still be difficult, which can prevent overall success. 
Therefore, complex sequential decision-making tasks usually 
require a significant amount of demonstrations to learn 
a reward function \cite{wulfmeier2016incorporating}.
 (2) \textit{\textbf{Burden of Human Interaction}} - Especially 
 in the case of human experts, constant human robot interaction 
 is very costly and should be minimized. Unfortunately, methods 
 like online imitation learning approaches often assume that the
 expert is always  providing demonstrations during the entire 
 learning process \cite{dagger}. While this may be reasonable 
for some problems, it maybe too impractical for many other applications. 
(3) \textit{\textbf{Data Redundancy}} - A full demonstration might not 
be needed for a learning agent equipped with a partial model. Given a 
small number of expert demonstrations, the learning agent may already 
know how to perform parts of the task successfully while struggling 
only in certain situations. In this case, it is more efficient to 
know where the agent fails and provide specific demonstrations for 
the part where the agent fails. 

We make the observation that human experts can provide high-level
feedback in addition to providing demonstrations for the task of
Inverse Reinforcement Learning (IRL). For example, in order 
to teach a complex task consisting of multiple decision-making steps, 
a common strategy of humans is to dissect the task into several 
smaller and easier subtasks \cite{narvekar2016source}
and then convey the strategy for each 
of the subtasks (see Figure~\ref{our_approach} for an example). 
It is reasonable that by incorporating this kind of divide-and-conquer 
high-level strategy coming from human's perception of the task, IRL 
can be more efficient by focusing on subtasks 
specified by human. In addition, by dividing a complex task into 
several subtasks using human's perception, it will be easier for
humans to evaluate the performance of the current agent. Since the
agent may already be able to perform well on some of the subtasks,
a human expert only needs to provide feedback on subtasks that the
agent struggles with.

We propose a Human-Interactive Inverse Reinforcement Learning 
(HI-IRL) approach that makes better use of human involvement 
by using \textit{structured} interaction. Although it requires 
more information from the human expert in the form of subgoal 
states, we demonstrate that this additional information significantly
reduces the required number of demonstrations needed to learn a task. 
Specifically, the human expert will provide critical subgoals (strategic information)
the agent should achieve in order to reach the overall goal. Thus, 
the overall task is more "structured" and consists of a set of subtasks. We
show that by using our sample efficient HI-IRL method, we can achieve
expert-level performance with significantly fewer human demonstrations than
several baseline IRL models. Additionally, we notice that the failure experience
obtained by the agent may also be helpful to learn the reward function since the
failure experience tells the agent of \textit{what not to do}. We leverage learning 
from failure experience to improve reward function inference.

\section{Related Work}
\label{ref:rel}

\textbf{Inverse Reinforcement Learning (IRL)}. IRL
is a method that infers a reward function given a set of expert demonstrations 
\cite{ng2000algorithms,abbeel2004apprenticeship}. One of the key
assumptions of IRL is that the observed behavior is optimal 
(maximizes the sum of rewards). Maximum entropy inverse reinforcement
learning \cite{ziebart2008maximum} employs the principle of maximum
entropy to learn a reward function that maximizes the posterior 
probability of expert trajectories. Though \cite{ziebart2008maximum} 
relaxes the optimality constraints, it cannot handle significantly 
suboptimal demonstrations. \cite{ziebart2008maximum} also does not consider
the redundancy of demonstrations. In our case, since we have both 
agent's failure experience as defined later and expert's demonstrations, we 
can leverage the failure experience to improve the current reward.
By using human feedback interactively in the training, our method aims to ultimately improve the reward inference process. By interacting with the human only when needed, we are also able to reduce the amount
of human involvement (\emph{i.e.,} redundant 
demonstration data).

\textbf{Human-in-the-Loop IRL}. Leveraging different types of human input during training has been previously shown to improve performance accuracy and learning efficiency. In \cite{hadfield2016cooperative}, the human and robot collaborate with each 
other to maximize the human's reward. Yet, \cite{hadfield2016cooperative} assumes that the underlying
reward function for every state is visible for the human, which may not be practical for many RL problems. One reason for this is that the human usually knows what action to take under a specific state, but it is hard to infer the value function of states as it triggers another IRL
problem. In \cite{odom2016active}, agents constantly seek advice from a human for clustered states, 
and so the learned reward gradually improves. However, creating 
the state clusters and give general advice for particular clusters 
is itself a demanding task for the human, since the states within
a cluster may not have the same optimal policy and the human has to 
tradeoff to make a decision. The work of \cite{2017arXiv170505427A} studied the safety of AI by giving human feedback when the agent is
performing sub-optimally, the method can reduce the amount of human 
involvement to learn a safe policy. However, the problem studied
is different from ours since we focus on improving IRL performance 
on complex sequential decision-making tasks instead of AI safety. 
As a human-in-the-loop imitation learning algorithm, DAGGER~\cite{dagger}
has proven to be effective in reducing the covariate shift problem 
in imitation learning. However, \cite{dagger} does not explicitly
learns a reward function and requires constant online interaction.

\textbf{Hierarchical IRL}. Hierarchical reinforcement learning 
\cite{hrl} was proved to be 
effective in learning to perform challenging tasks with sparse 
feedback by learning to optimize different levels of temporal 
reward functions. Hierarchical IRL
\cite{DBLP:journals/corr/KrishnanGLMPG16} was recently proposed
to learn the reward function for complex tasks with delayed 
feedback. The work of \cite{DBLP:journals/corr/KrishnanGLMPG16}
shows that by segmenting complex tasks into a sequence of 
subtasks with shorter horizons, it is possible to obtain 
optimal policy more efficiently. However, since
\cite{DBLP:journals/corr/KrishnanGLMPG16} does not get 
expert feedback during learning, and does not explicitly 
leverages partial demonstrations, it may still involve redundant
demonstrations.

\textbf{Learning from Failure}. 
Traditional IRL assumes the demonstrations by
experts are optimal in the sense that it optimizes the sum of reward
\cite{ng2000algorithms,ziebart2008maximum,levine2011nonlinear}. Recently, 
learning from failure experience has been proven to be beneficial with properly 
defined objective functions \cite{shiarlis2016inverse,lee2016inverse}. Inspired by \cite{shiarlis2016inverse}, we complement the human-in-the-loop training process with learning from failure experience experienced by agents, as we find it to improve reward function inference.

\section{Background}

\textbf{Maximum Entropy IRL.}
IRL typically formalizes the underlying 
decision-making problem as a \textit{Markov Decision Process} (MDP). 
An MDP can be defined as $\mathcal{M} = \{\mathcal{S}, \mathcal{A}, 
\mathcal{T}, r\}$, where $\mathcal{S}$ denotes the state space, 
$\mathcal{A}$ denotes the action space, $\mathcal{T}$ denotes the 
state transition matrix, and $r$ is the 
reward function. Given an MDP, an optimal policy $\pi^{*}$ is defined
as one that maximizes the expected cumulative reward. A discount 
factor $\gamma$ is usually considered to discount future rewards.

In IRL, the goal is to infer the reward
function given expert demonstrations $\mathcal{D} = \{d_0, d_1, \cdots,
d_N\}$, where each demonstration consists of state
action pairs $d_i = \{s_{i0}, a_{i0}, s_{i1}, a_{i1}, \cdots, s_{ik}, a_{ik}\}$.
The reward function is usually defined to be linear in the state features: 
$r = \theta^T\phi(s) = \theta^Tf_s$, where $\theta$ is the parameter of
the reward function, $\phi$ is a feature extractor, and $f_s$ is the 
extracted state feature for state $s$. In maximum entropy IRL, 
the learner tries to match the feature expectation to that of
expert demonstrations, while maximizing the entropy of the expert 
demonstrations. The optimization problem is defined as,
\begin{equation}
\theta^{*} = \arg\max_{\theta}-\sum_{d_i}P(d_i|\theta)\log{(P(d_i|\theta))},
\label{equ1}
\end{equation}
subject to the constraint of feature matching and being a probability
distribution,
\begin{equation}
\sum_{d_i}P(d_i|\theta)f_{d_i} = \tilde{f}^{\mathcal{D}},
\end{equation}
\begin{equation}
\sum_{d_i}P(d_i|\theta) = 1 \textrm{ and } P(d_i|\theta) \geq 0,  \forall i.
\end{equation}

The expert's feature expectation can be written as 
\begin{equation}
\tilde{f}^{\mathcal{D}} = \frac{1}{N}\sum_{d_i \in \mathcal{D}}\sum_{t=0}
^{k}f_{it}. \label{eq:expert_feature_expectation}
\end{equation}
Following current reward function $\theta$, the policy $\pi$ can 
be inferred via value iteration for low dimensional finite state 
problems. Then following $\pi$, and given initial state visitation
frequency $D_{s,0} = P(S_0=s)$ calculated from $\mathcal{D}$, 
the state visitation frequency at time step $t$ can be calculated as,
\begin{equation}
D_{s_i, t} = \sum_{k=0}^{|\mathcal{S}|}\sum_{j=0}^{|\mathcal{A}|}D_{s_k,t-1}\pi(s_k,a_{k,j})
\mathcal{T}(s_k,a_{k,j},s_i).
\end{equation}
Here $\pi(s_k,a_{k,j})$ is the probability of taking action $a_{k,j}$ when the 
agent is at state $s_k$, and $\mathcal{T}(s_k,a_{k,j}, s_i)$ is the
probability of transiting to state $s_i$ when the agent is at state $s_k$ and
taking action $a_{k,j}$. The summed state visitation frequency for each state is then
$D_{s_i} = \sum_{t}D_{s_i,t}$. The feature expectation following
current policy $\pi$ can be expressed as
\begin{equation}
f^{\pi} = \sum_{d_i}P(d_i|\theta)f_{d_i} = \sum_{s_i \in 
\mathcal{S}}D_{s_i}f_{s_i}. \label{eq:policy_feature_expectation}
\end{equation}

The above optimization problem in~\ref{equ1} can be transformed to the following
optimization problem \cite{ziebart2008maximum},
\begin{equation}
\begin{split}
\theta^{*} & = \arg\max_{\theta}P(\mathcal{D}|\theta) \\
& \propto \arg\max_{\theta}\exp{\{\sum_{s_i\in\mathcal{D}}\theta^T\phi(s_i)\}} \\
& = \arg\max_{\theta}\exp{\{\sum_{s_i\in\mathcal{D}}\theta^Tf_{s_i}\}}. \label{eq1}
\end{split}
\end{equation}

Optimizing Eq.~\ref{eq1} can be done via gradient descent on 
negative log-likelihood with the gradient defined by
\begin{equation}
\nabla_{\theta} =f^{\pi} - \tilde{f}^{\mathcal{D}}.
\label{optimize}
\end{equation}

\textbf{Maximum Entropy Deep IRL}.
Standard maximum entropy IRL uses a linear function to map state feature to 
reward value: $r = \theta^T f$. 
As neural networks have demonstrated excellent
performance in visual recognition and feature learning~\cite{krizhevsky2012imagenet}, 
it is reasonable that neural network-based reward
mapping function will be more powerful in complex state
space case, and can handle raw visual states which may be
challenging for linear reward function. 
The reward function is defined as $r = g(
\theta, f)$, where $r$ is the reward value for state
feature $f$, and $\theta$ is the neural network parameters. 
In the linear reward function case, the gradient of the loss
function with respect to the parameters is defined as,
\begin{equation}\begin{split}
\nabla_{\theta}L & = \nabla_{r}L\cdot\nabla_{\theta}r \\
& = \nabla_{r}L \cdot f.
\end{split}\end{equation}
From equation~\ref{optimize}, we know that $\nabla_{\theta}L = f^{\pi} - 
\tilde{f}^{\mathcal{D}}$, which can be expressed as,
\begin{equation}
f^{\pi} - \tilde{f}^{\mathcal{D}} = f(D_{f}^{\pi} - \tilde{D}_{f}^{\mathcal{D}}),
\end{equation}
where $f$ is the feature of a particular state, $D_{f}^{\pi}$ 
is the agent visitation frequency of this state, and 
$\tilde{D}_{f}^{\mathcal{D}}$ is the expert visitation 
frequency of this state. When deep neural network is
used to represent the reward function, the gradient of 
the loss function with respect to the parameters can 
be expressed as,

\begin{equation}\begin{split}
\nabla_{\theta}L & = \nabla_{r}L\cdot\nabla_{\theta}r \\
& = \nabla_{r}L \cdot \nabla_{\theta}g \\ 
& = (D_{f}^{\pi} - \tilde{D}_{f}^{\mathcal{D}}) \cdot\nabla_{\theta}g.
\end{split}\end{equation}

\textbf{IRL from Failure}.
While maximum entropy IRL tries to match the expected feature counts 
of the agent's trajectory with the feature counts of expert demonstration, 
it is reasonable to keep the expected feature counts following current 
learned reward different from that of failure experience. The 
learning from failure algorithm proposed in \cite{shiarlis2016inverse}
demonstrates the possibility of incorporating failure experience 
to improve IRL. Given both successful demonstrations $\mathcal{D}$ and 
failure experience $\mathcal{F}$, we define linear 
reward function parameter $\theta_d$ and $\theta_f$ for 
reward function learned from $\mathcal{D}$ and $\mathcal{F}$ respectively. 
The goal is to maximize the probability of successful demonstrations,
and match the feature expectation of successful demonstrations,
while maximizing the feature expectation difference with failure
experiences. In \cite{shiarlis2016inverse}, the 
optimization problem is defined as following,
\begin{equation}
\begin{split}
\max_{\pi, w, z} \quad & H(\mathcal{D}) + wz - \frac{\lambda}{2}\|w\|^2, \\
\textrm{subject to:} \quad & \tilde{f}^{\mathcal{D}} = f^{\pi}, \\
\quad &  f^{\pi} - \tilde{f}^{\mathcal{F}} = z, \\
\quad & \sum_{a}\pi(s,a) = 1 \quad \forall s \in \mathcal{S}, \\
\quad & \pi(s,a) \geq 0 \quad \forall a \in \mathcal{A}, \\
\end{split}
\label{eq3}
\end{equation}
where $H(\mathcal{D})$ is the causal entropy of the successful 
demonstrations $\mathcal{D}$, and is defined as,
\begin{equation}
H(\mathcal{D}) = -\sum_{t}\sum_{s_{1:t}\in\mathcal{S}, 
a_{1:t}\in\mathcal{A}}P(a_{1:t},s_{1:t})
\log{(P(a_t|s_t))},
\end{equation}
where $P(a_t|s_t) = \pi(s_t,a_t)$ is the policy, 
and 
\begin{equation}
    P(a_{1:t},s_{1:t}) = P(s_{1:t-1},
a_{1:t-1})\mathcal{T}(s_{t-1},a_{t-1},s_t)\pi(s_t,a_t)
\end{equation} 
is the 
probability of trajectory from time $1$ to time $t$. In Eq.~\ref{eq3},
$w$ is the Lagrange multiplier of $z$, which is a variable
representing the difference between the feature expectation of failure
experiences and the feature expectation following current policy $\pi$.
The Lagrangian of Eq.~\ref{eq3} gives the following loss function, 
\begin{equation}
\begin{split}
\mathcal{L}(\pi, w, z, \theta_d, \theta_f) = & 
H(\mathcal{D}) + wz - \frac{\lambda}{2}
\|w\|^2 \\
& + \theta_d(f^{\pi} - \tilde{f}^{\mathcal{D}}) \\
& + \theta_f(f^{\pi} - \tilde{f}^{\mathcal{F}} - z). \\
\end{split}
\end{equation}
Following the optimization in \cite{shiarlis2016inverse}, the optimization step
update for $\theta_d$ and $\theta_f$ is,
\begin{equation}
\begin{split}
    \theta_d & = \theta_d - \alpha(f^{\pi} - \tilde{f}^{\mathcal{D}}), \\
    \theta_f & = \frac{(f^{\pi} - \tilde{f}^{\mathcal{F}})}{\lambda},
\end{split}
\label{eq15}
\end{equation}
where $\alpha$ is the learning rate for $\theta_d$ and $\lambda$
is a learning rate for $\theta_f$ which is annealed throughout 
the learning. More details of the learning from failure approach can be
found in \cite{shiarlis2016inverse}.

\begin{algorithm}[t]
\caption{Deep IRL from Failure (\textsc{IRLFF}) 
}
\begin{algorithmic}
\STATE \textbf{Require:} Failure experience $\mathcal{F}$, expert
demonstration $\mathcal{D}$ \STATE \textbf{Require:} State Transition
Matrix $\mathcal{T}$, $\alpha$, $\alpha_{\lambda}$, $\lambda$,
$\theta_d^{t}$, $\lambda_{min}$, all feature input 
$f$, where $\theta_d^t$ is a deep neural
network
\STATE \textbf{Return:} Updated reward function $\theta_d,\theta_f$
\STATE \textbf{Start:}
\STATE \hspace{1cm} $\tilde{f}^{\mathcal{D}}$ = \textsc{FeatureCount}$(\mathcal{D})$ \textit{(Eq.~\ref{eq:expert_feature_expectation} with $\mathcal{D}
=\mathcal{D}$)} 
\STATE \hspace{1cm} $\tilde{f}^{\mathcal{F}}$ = \textsc{FeatureCount}$(\mathcal{F})$ \textit{(Eq.~\ref{eq:expert_feature_expectation} with $\mathcal{D}
=\mathcal{F}$)}
\STATE \hspace{1cm} $P_0^{\mathcal{D}}$= initialStateDistribution$(\mathcal{D})$
\STATE \hspace{1cm} $P_0^{\mathcal{F}}$ = initialStateDistribution$(\mathcal{F})$
\STATE \hspace{1cm} $\theta_f = \textbf{0}$
\STATE \hspace{1cm} $\theta_d = $ $\theta_d^{t}$
\STATE \hspace{1cm} \textbf{Repeat:}
\STATE \hspace{2cm} $r = g(\theta_d,f) + 
\theta_f\cdot g_{fc1}(\theta_d,f)$
\STATE \hspace{2cm} $\pi$ = \textsc{SoftValueIteration}($r$)
\STATE \hspace{2cm} $f^{\pi}_{\mathcal{F}}$ =
\textsc{FeatureExpectation}($P_0^{\mathcal{F}}, \pi, \mathcal{T}$)
\STATE \hspace{2cm} $f^{\pi}_{\mathcal{D}}$ =
\textsc{FeatureExpectation}($P_0^{\mathcal{D}}, \pi, \mathcal{T}$)
\STATE \hspace{2cm} $\theta_d = \theta_d - \alpha(D^{\pi}_{f}-\tilde{D}^{\mathcal{D}}_f)\cdot \nabla_{\theta_d}g$
\STATE \hspace{2cm} $\theta_f$ calculated according to Eq.~\ref{eq18}
\STATE \hspace{2cm} \textbf{if} $\lambda > \lambda_{min}$: 
\STATE \hspace{3cm} $\lambda = \alpha_{\lambda}\lambda$
\STATE \hspace{1cm} \textbf{until convergence}
\end{algorithmic}
\label{alg2}
\end{algorithm}

\section{Human-Interactive Inverse Reinforcement Learning (HI-IRL)}\label{sechiirl}

We propose Human-Interactive Inverse Reinforcement Learning (HI-IRL)
to make more efficient use of human participation beyond simply providing demonstrations. Different from approaches such as \cite{ziebart2008maximum},
we require more human-agent interactions during the learning process
by allowing the agent try out subtasks defined by a human and letting the 
human provide further demonstrations on subtasks if the agent struggles (we provide formal definition of ``struggle'' later in this section). Different from 
approaches such as DAGGER~\cite{dagger}, humans do not need to constantly 
provide entire demonstrations; instead demonstrations are obtained 
only when required by the agent. There indeed can be other forms of human interaction
when the agent struggles, some of which are compared to as baselines in the experiments. For example, the human may continue to provide the entire
demonstrations when the agent struggles, similar to the approach in 
\cite{dagger}. However, we find this method of interaction to be less effective. A second possibility is to simply let the agent
try the same task repeatedly, until it happens 
to finish the task. Then, the successful trajectory that the agent experienced can be used as 
human demonstration. However, this approach is limited in scenarios with large state spaces. In addition to being highly inefficient, even if the 
agent reaches the goal, the trajectory that the agent traveled may not be an optimal or a desired trajectory. In contrary, we show that our method of structuring the interaction enables better efficiency on complex tasks. Next, we first describe our method, HI-IRL, and then give a demonstration of the optimality of our subgoal selection strategy.


\begin{figure*}[t]
    \centering
    \includegraphics[width=0.20\textwidth, height=0.20\textwidth]{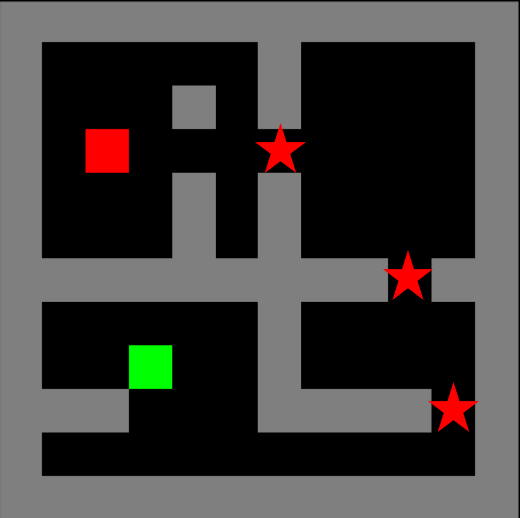}
    \includegraphics[width=0.20\textwidth, height=0.20\textwidth]{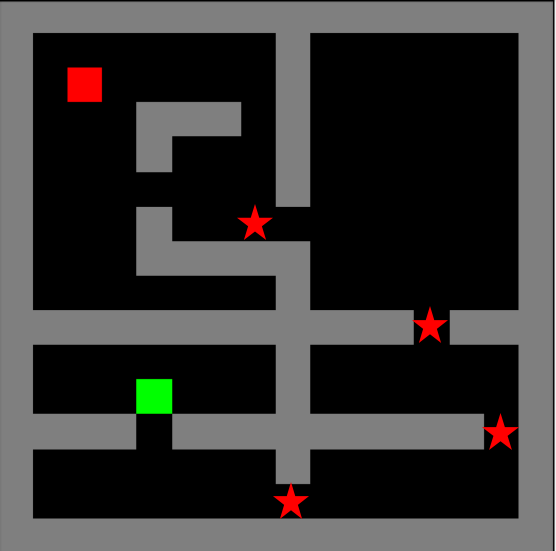}
    \includegraphics[width=0.20\textwidth, height=0.20\textwidth]{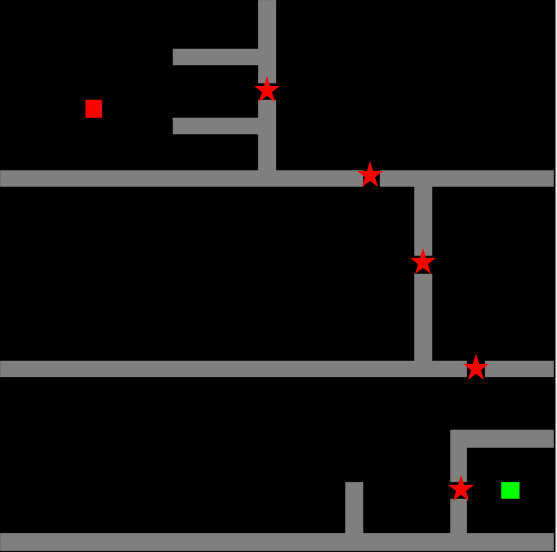}
    \includegraphics[width=0.20\textwidth, height=0.20\textwidth]{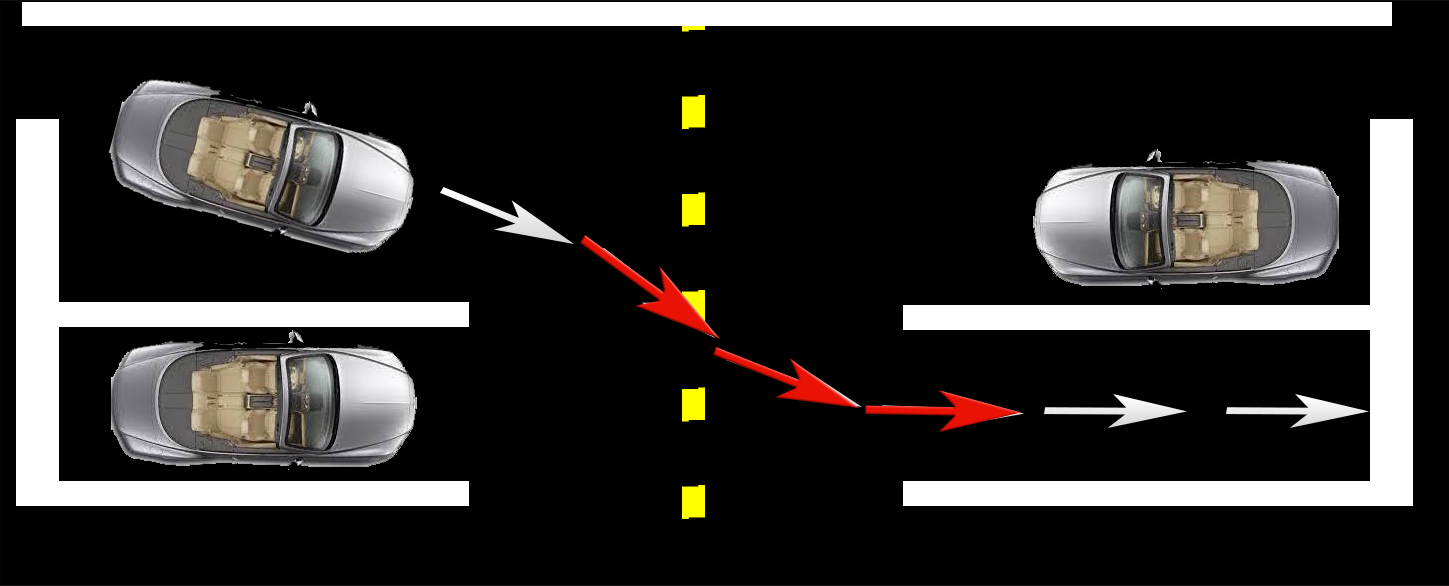}
    \caption{Subgoals specified in 12x12, 16x16, 32x32 grid world 
    environment, and car parking environment. In the grid world 
    environment, states are defined as the grid position the agent
    is current at (specified by red box), goals are represented by
    green box, and subgoals are indicated by red stars. In the car
    parking environment, states are defined
    as the car global coordinate as well as the orientation of the car,
    which can be represented by an arrow. The subgoals in the car parking environment is specified by a set of red arrows. }
    \label{fig:subgoal}
\end{figure*}

\subsection{HI-IRL} \label{subsec:hiirl}

\textbf{Step 1: Human expert provides several full demonstrations and 
define subgoals}. Given a task consisting of multiple decision making steps,
the human expert
will first provide $N$ full demonstrations $\mathcal{D}=\{d_0, \cdots, d_N\}$ 
completing the entire task. The number of demonstrations in $\mathcal{D}$ 
can be relatively small, for example, 1 or 2 demonstrations to learn an 
initial reward function. The human expert will then dissect the entire task into 
several parts by indicating critical subgoal states where the agent must go 
through in order to achieve the overall task. For example, in an indoor
navigation task, the agent tries to find a way from one room to anther, the state
when the agent is at the exit between the two rooms is a critical subgoal
state. While trajectories with different starting position in the first room 
and different goal position in the second room varies, they all need
to go through the critical state corresponding to the exit. 

We denote these critical subgoal states as $\mathcal{S}_{sub}$. One typical 
characteristics of these subgoal states is that the probability of
any expert trajectories to include them will be close to 1,
\begin{equation}
    P(s_i \in d_j) \approx 1, \forall s_i\in\mathcal{S}_{sub}
    \text{ and }\forall d_j\in \mathcal{D}.
\end{equation}
The reason why it may not be 1 is to allow cases where there are 
multiple states functioning very similar as subgoal states. For
instance, there are multiple exits from one room to another in the
indoor navigation example. In this case, the probability of any 
expert trajectories to include any one of these states will be 1. 

Given these subgoal states $\mathcal{S}_{sub}$, any trajectory
$\xi=\{s_0,\cdots,s_k\}$ can be dissected into several subtasks 
$T_{sub} = \{d_1, d_2, \cdots, d_m\}$, where $m$ is the number of
subtasks within this trajectory $\xi$ 
, and concatenating these subtasks
together will get the original trajectory $\xi$. The starting state
and end state of each of these subtasks except $d_1$ and $d_m$ belong
to $\mathcal{S}_{sub}$. The end state and starting state of $d_1$ and
$d_m$, respectively, belong to $\mathcal{S}_{sub}$. A more formal 
definition of trajectory dissection is to consider all possible
trajectories from a chosen start state to goal state as a set 
$\Xi=\{\xi_1, \cdots, \xi_x\}$, and subgoal states are defined by,
\begin{equation}
    \mathcal{S}_{sub} = \bigcap_{i=1}^x{\xi_i}.
\end{equation}

\textbf{Step 2: Agents tries the defined subtasks}. Starting
from a randomly selected starting state $s_r$, the agent will 
be required to reach each of the subgoals sequentially towards the
ultimate state $s_{goal}$. This means that given the optimal path 
from the agent's current state $s_r$ to the goal state $s_{goal}$: 
$\xi_{s_r\rightarrow s_{goal}} =\{s_r, \cdots, s_{sub1},\cdots, \\
s_{sub2}, \cdots, s_{subk}, \cdots, s_{goal}\}$ where the agent is expected to reach subgoal states along the path from $s_{sub1}$ to $s_{subk}$
sequentially. If the agent successfully
arrives to subgoal $s_{subi} $ within $step_{min,s_{subi}}+step_{thr}$, the agent will
be required to reach the next subgoal $s_{sub(i+1)}$ starting from current state $s_{subi}$. Here, $step_{min,s_{subi}}$ is the minimum steps required to reach 
$s_{subi}$ from the start state $s_r$, and $step_{thr}$ is the extra threshold
steps to allow some exploration.

\begin{algorithm}[t]
\caption{Human-Interactive Inverse Reinforcement Learning (HI-IRL)}
\begin{algorithmic} 
\STATE \textbf{Require:} Set of initial demonstrations $d_0$, $T$, State Transition Matrix $\mathcal{T}$, $\theta^0$, all state raw feature $f$, and human $\mathcal{H}$.
\STATE \textbf{Return:} Reward function $\theta^{T+1}$
\STATE \textbf{Define:} $\mathcal{D}$: positive demonstrations; $\mathcal{F}$: failure experience; $\mathcal{E}$: agent experience;
$\mathcal{S}_{sub}$: set of subgoal states
\STATE \textbf{Start:}
\STATE \hspace{0.5cm}$\mathcal{S}_{sub}$ = specify\_subgoals($\mathcal{H}$)
\STATE \hspace{0.5cm}$\mathcal{D} = d_0$;
\STATE \hspace{0.5cm}$\theta^{1}$ = MaxEntIRL($\mathcal{D}$, $\theta^{0}$)
\STATE \hspace{0.5cm} \textbf{for} t $\in 1,2,..., T$
\STATE \hspace{1cm} $\mathcal{E} = $ \textsc{Rollout}($\theta^{t}$, $\mathcal{S}_{sub}$)
\STATE \hspace{1cm} \textbf{for} $e$ in $\mathcal{E}$
\STATE \hspace{1.5cm} $\mathcal{F}, \mathcal{D} = $ \textsc{UpdateDemo}($e$, $\theta^{t}$, $\mathcal{D}$)
\STATE \hspace{1cm} $\theta_d^{t+1}, \theta_f^{t+1}$ = \textsc{IRLFF}($\mathcal{F}, \mathcal{D}$, 
$\mathcal{T}$, $\theta_d^{t}$, $f$) \textit{(Alg.~\ref{alg2})}
\STATE \hspace{1cm} $\theta^{t+1} = (\theta_d^{t+1}, \theta_f^{t+1})$
\end{algorithmic}
\label{alg1}
\end{algorithm}

\textbf{Step 3: Human provides further demonstrations if needed}. Depending
on the performance of the agent on the subtasks, if the agent successfully
finished all subtasks, then the human expert will not provide further
demonstrations. The human expert will only provide demonstrations on 
subtasks that the agent struggles. For example, if the agent is not able to complete a subtask ending in subgoal $s_{subi}$, then human will provide
further demonstrations on this subtask. Since these additional demonstrations
may not be complete demonstrations starting from the very beginning state
to the ultimate goal state, we refer to these demonstrations as \textit{partial
demonstrations}. The initial demonstrations mentioned in step 1 are referred as \textit{full demonstrations}. This intuitive interaction scenario is formally defined below.

Suppose the agent is
given a subtask to go from state $s_i$ to state $s_j$. The minimum number
of steps to travel from $s_i$ to $s_j$ is $step_{min, s_i\rightarrow s_j}$, and to
allow some level of exploration, the agent will be given extra $step_{thr}$ steps
to reach $s_j$. The value $step_{thr}$ depends on the difficulty of specific
task, if the task is fairly difficult, we set it to a high value, otherwise,
we set it to a low value. In our approach this value can be regarded as a hyper-parameter
that needs to be tuned. Struggling is defined as the scenario where the agent is not able to reach $s_j$ within $step_{thr}+
step_{min, s_i\rightarrow s_j}$. Here, the human will provide further demonstrations on this particular
task (from $s_i$ to $s_j$). 

\textbf{Step 4: Learning reward function from both failure experiences 
and expert demonstrations}. When the agent fails to finish 
some subtasks, it gains failure experiences, denoted as $\mathcal{F}$. These demonstrations are not given
by human, but instead by the learning agent itself. 
The expert's further demonstrations are denoted
as $\mathcal{D}$, which already includes the initial full demonstrations.
Since learning from failure approaches~\cite{shiarlis2016inverse} generally focus on the linear
reward function case, we propose to use a deep neural network to extract
features from raw states, and then use a linear reward function to get
reward value from these extracted features. 

Our deep neural network reward function takes in input in the form of
raw states (\emph{i.e.,} images) and process it with 
three convolutional layers with each one followed by batch normalization
layers and ReLU activation. Two fully connected layers are
followed to output the final reward value. The last layer outputs a scalar
value which will be used as the reward value corresponding to $\theta_d$ in Eq.~\ref{eq15}. The second last layer output vector will be used to
calculate $\theta_f$ in Eq.~\ref{eq15}. If we denote the network
parameters as $\theta_d=\{conv,bn,ReLU,FC_1, FC_2\}$, the network input as
$f$, and the network function as $r_d = g(\theta_d, f)$,
then we have
\begin{equation}
\centering
    \begin{split}
    FC_{1,out} &= g(conv,bn,ReLU,FC_1,f) \doteq g_{fc1}(\theta_d,f)\\
    \theta_f & = \frac{FC_{1,out}^{\pi} - \widetilde{FC}_{1,out}^{\mathcal{F}}}{\lambda} \\
    \end{split}
    \label{eq18}
\end{equation}
Here $\theta_d$ will be the neural network and $\theta_f$ will be a vector
of the same size as $FC_{1,out}$, $FC_{1,out}^{\pi}$ is the feature
expectation following the current policy $\pi$, and
$\widetilde{FC}_{1,out}^{\mathcal{F}}$ is the feature expectation of failure
experience $\mathcal{F}$. The final reward function will be $r=g(\theta_d,f)
+\theta_f\cdot g_{fc1}(\theta_d,f)$. 
The detailed learning from both failure
experience and expert demonstration algorithm is described in 
Algorithm~\ref{alg1}.

\subsection{Optimality of Subgoal Selection}
In HI-IRL, the human will specify critical subgoal 
states $\mathcal{S}_{sub}$ which have a very high 
probability to be included in any expert demonstrations, and other
non-critical states will have relatively lower probability to be
included in any expert demonstrations. Define $\mathcal{S}_{nc}\doteq
\mathcal{S}\setminus\mathcal{S}_{sub}$ as all states except human
defined subgoal states. Given two trajectories $\xi_{1}=\{s_{1,0}, s_{1,1},\cdots, s_{1,k}\}$ and $\xi_2=\{s_{2,0}, s_{2,1},\cdots, s_{2,k}\}$, where $s_{1,i}=s_{2,i}, \forall i \in \{0,\cdots,
k-1\}$, and $s_{1,k}\in \mathcal{S}_{sub}$ 
and $s_{2,k}\in \mathcal{S}_{nc}$, intuitively, $\xi_1$ will be
favored over $\xi_2$, 
\begin{equation}\begin{split}
    P(\xi_1) & > P(\xi_2) \\
    \Rightarrow  \exp{\sum_{i=1}^k{r(s_{1,i})}} & > \exp{\sum_{i=1}^k{r(s_{2,i})}}, \\
    \Rightarrow {r(s_{1,k})} & > {r(s_{2,k})},\\
\end{split}
\label{eq19}
\end{equation}
which means that critical subgoal states will have higher reward
than non-subgoal states around them. In the linear reward function
case, the reward function parameter $\theta$ is optimized when,
\begin{equation}
    \tilde{f}^{\mathcal{D}} = \sum_{i=1}^{|\mathcal{S}|}D_{s_i}f_{s_i},
\end{equation}
which means the final policy will favor states that appear more
times in expert demonstrations $\mathcal{D}$ in order to match 
the feature expectation of $\mathcal{D}$. Given two states $s_1$ and
$s_2$, and define $p(s_1,\mathcal{D})$ as the frequency of $s_1$ appears
in $\mathcal{D}$, the same for $s_2$, and suppose $p(s_1,\mathcal{D})>p(
s_2, \mathcal{D})$, then we have,
\begin{equation}
\begin{split}
    D_{s_1} & > D_{s_2} \\
    \Rightarrow P(\xi_1|s_1\in\xi_1) & > P(\xi_2|s_2\in\xi_2), \\
    \end{split}
\end{equation}
where $\xi_1$ and $\xi_2$ are two trajectories, where all other
states are same, except that $\xi_1$ contains $s_1$ while $\xi_2$
contains $s_2$. Given Eq.~\ref{eq19}, we know that $r(s_1)>r(s_2)$,
which means states that appear more times in expert demonstrations
will typically have higher rewards. Therefore, in order to make sure
those critical states have higher rewards, we must increase the 
demonstrations around them. By letting human specify these critical
states, and providing extra demonstrations if the agent struggles,
we ensure that these states receive more attention during demonstration
collection, which leads to better reward function learning.

\begin{figure*}[t]
    \centering
    \begin{subfigure}{.24\textwidth}
        \centering
        \includegraphics[width=\textwidth]{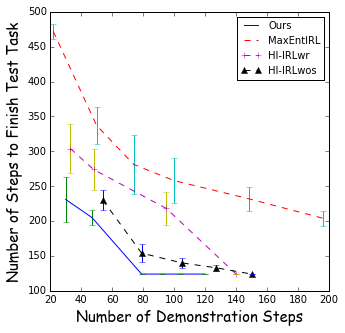}
        \caption{}
    \end{subfigure}%
    \begin{subfigure}{.24\textwidth}
        \centering
        \includegraphics[width=\textwidth]{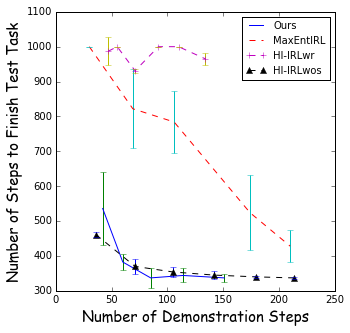}
        \caption{}
    \end{subfigure}
    \begin{subfigure}{.24\textwidth}
        \centering
        \includegraphics[width=\textwidth]{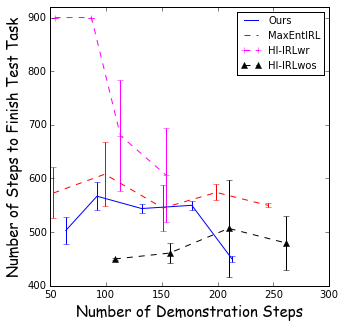}
        \caption{}
    \end{subfigure}%
    \begin{subfigure}{.24\textwidth}
        \centering
        \includegraphics[width=\textwidth]{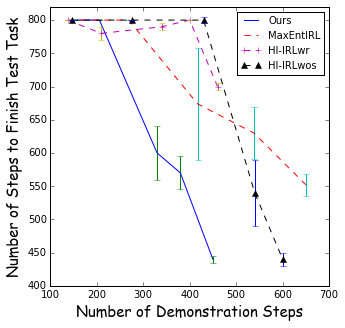}
        \caption{}
    \end{subfigure}%
    \caption{Number of demonstration steps VS number of steps used to complete
    the same test tasks curve. (a): 12x12 Grid-world; (b) 16x16 Grid-world; (c)
    32x32 Grid-world. (d) Car parking environment.}
    \label{curve1}
\end{figure*}

\section{Experiments}

We designed the experiment parts to demonstrate the key contributions 
of our proposed HI-IRL method. \textbf{First}, 
we demonstrate that by leveraging human interaction in inverse reinforcement
learning, we obtain better data efficiency than traditional inverse reinforcement
learning approach that trains on offline collected data
(the standard maximum entropy IRL method). \textbf{Second},
we provide a better human interaction strategy where the burden
on human can be reduced compared with existing methods such as
\cite{dagger}. \textbf{Third}, we demonstrate that by 
carefully selecting the key subgoals,
it achieves better reward function learning than random selection of
subgoals. The experimental environments are designed to be complex
sequential decision making process with critical subgoal states that
the agent must go through in order to complete the overall task. 

\textbf{Baselines.} In order to show the key contributions of
our HI-IRL method, we compare our algorithm with 
(1) maximum entropy IRL (here after 
denoted as \texttt{\textbf{MaxEntIRL}}); 
(2) human interactive IRL without specifying subgoals 
(here after denoted as \texttt{\textbf{HI-IRLwos}}),
which is similar to approach like \cite{dagger}; 
and (3) human interactive IRL with randomly selected subgoals (
here after denoted as \texttt{\textbf{HI-IRLwr}}). In human interactive IRL 
without specifying subgoals, the procedure is similar to our 
method, except that the agent will be required to complete 
entire task and human expert will provide full demonstrations
if the agent struggles. The purpose of comparing with 
\texttt{\textbf{MaxEntIRL}} is to show the benefits of interacting
with human during the learning process (our \textbf{first} 
contribution). While both
\texttt{\textbf{HI-IRLwos}} and \texttt{\textbf{HI-IRLwr}} have
human interaction, \texttt{\textbf{HI-IRLwos}} tries to provide
the entire demonstration again which contains redundancy and 
increases human burden; \texttt{\textbf{HI-IRLwr}} tries to provide
demonstrations for randomly selected subtasks, which fails to
emphasize on critical subgoal states, and may lead to ill reward
function learning. The purpose of comparing with \texttt{\textbf{HI-IRLwos}}
is to show the benefits of subgoal selection as it reduces human
burden to demonstrate entire task (our \textbf{second} 
contribution). The purpose of comparing with \texttt{\textbf{HI-IRLwr}}
is to show the benefits of selecting critical subgoals instead of
random subgoals (our \textbf{third} contribution).

We performed several sets of experiments in grid-world and car 
parking environments spanning 
different scales of state space. All environments
contain critical subgoal states that the agent \textit{must}
go through to complete the entire task. 
In all experiments, we use deep neural network to 
represent reward function. 
In the grid-world environment, the network is composed of
three layers of convolutional neural network with each
followed by a batch normalization layer and ReLU activation layer,
then two fully connected layers are followed to output the final
reward value. In the car parking environment, the network is similar to
the network in grid-world environment, except there are 2 
convolutional layers due to smaller input image size.

\textbf{Grid-world Environment}. The grid-world environment involves 
grid-world navigation where the agent is put in a place 
at the beginning and the task is to find a way to a target 
position. In this experiment, grid-world of different
scales of state space are used for evaluation. Specifically,
a 12x12, a 16x16, and a 32x32 grid-world environment are used. Regions in
the grid-world where there are obstacles are not counted towards
agent state.  

Since all four methods require some initial human 
demonstration to learn a reward function, a certain number of human 
demonstrations $\mathcal{D}$ are collected at the beginning. 
In both the gridworld environment and car parking environment, 
we have finite number of states and the optimal path from one state 
to another can be automatically solved by using the Dijkstra algorithm 
\cite{skiena1990dijkstra}. Therefore, we generate the demonstration 
automatically instead of getting them from real human. 
However, human expert will specify critical subgoal states 
$\mathcal{S}_{sub}$ to be used in our method. A set of
test starting state will be specified by human that is
different from the training data $\mathcal{D}$. Then
$\mathcal{D}$ is used to get the reward function following
\texttt{\textbf{MaxEntIRL}} method. One demonstration 
randomly sampled from $\mathcal{D}$
will be used for training initial reward function for our method,
\texttt{\textbf{HI-IRLwos}} method, and \texttt{\textbf{HI-IRLwr}} method. 
In \texttt{\textbf{HI-IRLwos}}, the agent 
will be required to start from a randomly selected
starting state, and find a way to the final target state, and human
will provide further demonstration if the agent struggles. 
In \texttt{\textbf{HI-IRLwr}}, randomly selected subgoals will be used to define 
subtasks, and the agent will try to complete these subtasks, 
and human will provide further demonstrations if needed. 
All four methods are trained with the same learning rate and number of iterations.
Different number of demonstrations are used to train reward function and then
evaluate on the same test task 5 times to get the mean value of test performance.

\textbf{Car-Parking Environment}.
Parking a car into a garage spot involves driving the car to 
a place near the slot, adjust the orientation of the car and
drive the car into the parking box without hitting obstacles. 
In this environment, it is critical that the car has to stop 
at a certain state near the parking slot to ensure that 
after adjusting the orientation, the car will not hit obstacles.
The car parking environment interface is shown in Figure~\ref{parking_car}.
The number of agent possible states is about 5k -- much larger
than the state space in the grid-world environment. 

At the beginning, human demonstrations and human specified
subgoals are collected. Then follow the same procedure as in
the grid-world environment, we obtained training results for all four
methods. The subgoals selected for each environment is visualized in
Figure~\ref{fig:subgoal}.

\subsection{Results and Analysis}

\textbf{Grid-world Environment}.
The number of demonstration steps versus number of steps used to 
complete the same test tasks curve is shown in Figure~\ref{curve1},
which includes the results for all four methods. The test task
is to set the agent at some initial states on the top left region
in the grid world, and then require the agent to travel to the same
destination as in training time. Since the goal
of our approach is to reduce the burden of human, for example,
the human will provide less demonstrations, the results indicate
that our method achieves better human interaction efficiency
and the agent learns to complete the same test task with less
but more informative demonstration from human. The reason why
the \texttt{\textbf{MaxEntIRL}} method works worse than the other three
methods is that there are much more training data to learn from
in this method. Therefore, it may require more iterations to train,
which is another burden of this method. The \texttt{\textbf{HI-IRLwr}}
method works in the 12-by-12 state size case, but does not work
in the 16-by-16 state size case. The reason is that the subgoals
are randomly selected, which means there is a probability that they
are selected to be near the critical subgoal states, achieving 
similar performance as our method. Our method uses slightly more
steps to complete the test task in the 32-by-32 grid-world at initial
training than \texttt{\textbf{HI-IRLwos}} method. However, as indicated in
the figure, we can use less steps of demonstrations but achieve
similar performance.

\textbf{Car-Parking Results}. The car-parking results 
include the number of demonstrations versus number of steps to complete
the same test tasks curve shown in Figure~\ref{curve1}.
Our method achieves near oracle performance with less demonstrations
from human than other baselines. 
Since this MDP contains much richer states (in total 5k states)
than previous MDPs, this experiment demonstrates that our method
has the abilility to generalize to large state space case.

\section{Conclusions and Remarks}

Motivated by the need to address challenges when 
learning complex sequential decision-making with an IRL framework, this paper presents a framework for leveraging structured 
interaction from a human during training. In addition to providing
demonstrations of the task to be performed by a learned agent, 
the method also leverages the human's high level perception 
about the task (in the form of subgoals) in order to improve 
learning. Specifically, humans can transfer their 
divide-and-conquer approach for problem solving to inverse 
reinforcement learning by providing segmentation of the current
task and a set of subtasks. Additional improvements are made by employing the agent's own
failure experience in addition to the human's demonstrations. Experiments on a discrete grid-world path-planning task and 
large state space car parking environment demonstrated how
subgoal supervision resulted in more efficient learning.  

For future work, we would like to apply HI-IRL for additional 
tasks with increasing complexity. Incorporating HI-IRL with a 
real-world robot experiment could further support its use in 
applications where input from a human is helpful but costly 
to acquire. In addition, it is also interesting to explore automatic
optimal task dissection to further reduce human burden.



\bibliographystyle{ACM-Reference-Format}  
\balance
\bibliography{sample-bibliography}  

\end{document}